\documentclass[pre,twocolumn,superscriptaddress,showpacs,preprintnumbers,
eqsecnum,amsmath,amssymb]{revtex4}

\usepackage{graphicx}

\begin{document}

\title{Experimental determination of the absorption strength in absorbing chaotic cavities}

\author{G. B\'aez}
\affiliation{\'Area de F\'isica Te\'orica y Materia Condensada, Universidad Aut\'onoma Metropolitana-Azcapotzalco, A. P. 21-267, 04000 M\'exico D. F., Mexico}

\author{M. Mart\'{\i}nez-Mares}
\affiliation{Departamento de F\'{\i}sica, Universidad Aut\'onoma
Metropolitana-Iztapalapa, A. P. 55-534, 09340 M\'exico D. F., Mexico}

\author{R. A. M\'endez-S\'anchez}
\affiliation{Instituto de Ciencias F\'{\i}sicas, Universidad Nacional Aut\'onoma de M\'exico, A. P. 48-3, 62210 Cuernavaca Mor., Mexico}

\date{\today}

\begin{abstract} 
Due to the experimental necessity we present a formula to determine the absorption strength by power losses inside a chaotic system (cavities, graphs, acoustic resonators, etc) when the antenna coupling, always present in experimental measurements, is taken into account. This is done by calculating the average of the absorption coefficient as a function of the absorption strength and the coupling of the antenna to the system, in the one channel case.
\end{abstract}


\maketitle


\section{Introduction}
\label{sec:intro}

A great interest on chaotic systems with internal losses, or absorption, is
due to the fact that they are good candidates to verify Random Matrix Theory 
(RMT). The effect of losses has given rise to many theoretical and experimental investigations on transport properties (for a review see Ref. \cite{Fyodorov2005}, see also \cite{interpolation,Hemmady2006,moises-pier,Dominguez,GMMMS}). In particular microwave \cite{Richter,Stoekmann} and acoustic resonators \cite{Schaadt}, microwave networks \cite{Hul2004} and elastic systems \cite{Morales} are  excellent tools for that purpuse.

One of the most important things in absorbing systems is to quantify the degree of the losses suffered in an experimental situation. Although in some experiments the absorption strength $\gamma$ can be partially controlled by attaching aditional antennas \cite{Barthelemy2005} or introducing absorber materials inside the system \cite{Hemmady2006}, the intrinsic absorption is not directly under experimental control. However, it can be tunned to fit the experimental data \cite{Mendez-Sanchez2003,Schanze2004,Kuhl2004}. In a single-mode port microwave resonator setup of Refs.  \cite{Mendez-Sanchez2003,Kuhl2004}, the one channel case,  $\gamma$ was determined as the one that reproduces the experimental value of the average of the reflection coefficient $\langle R\rangle$, which is a monotonically decreasing function of $\gamma$. A similar procedure was used in Ref. \cite{Schanze2004} but with the transmission coefficient $T$ in a two single-mode ports. In that case, $\gamma$ was chosen as the value in which the theoretical distribution of $T$ fits the experimental one. Also, the value of $\gamma$ was estimated from the fidelity \cite{Seligman}. 

The imperfect coupling of the antennas to the cavity has a similar effect as the absorption and unfortunately the last procedure can not discern what part of $\gamma$ is due to losses and what to the rejected part of the wave that never enter the cavity. In this sense an imperfect coupling mimics absorption and a procedure that gives the correct value of $\gamma$ is needed. Here we present a semi-analytical formula to calculate $\gamma$ which takes into account the coupling for the one channel case. This is helpful in experiments with microwave networks and one port cavities. Although the procedure is the same for higher number of channels, it is not possible to give a result due  to the absence of knowlegde of the probability distribution of the scattering matrix \cite{Fyodorov2005}.

In the next section we summarize the existing theory used to describe the scattering process through chaotic cavities with losses, taking the imperfect coupling to a single-mode port into account. Sect. \ref{sec:gamma} is devoted to calculate the absorption strength $\gamma$ in terms of the average of $R$ the coupling intensity. We conclude in Sect. \ref{sec:conclusions}.


\section{The ${\tilde S}$-matrix and its distribution}
\label{sec:distS}

In a single-mode port cavity in the presence of losses, the scattering matrix ${\tilde S}$ is a subunitary $1\times 1$ matrix that we parametrize as
\begin{equation}
{\tilde S} = \sqrt{R}\,e^{i\theta}.
\end{equation} 
Here $R$ is the reflection coefficient and $\theta$ is minus twice the phase shift due to the scattering process into the cavity. Those quantities are measured in actual experiments with microwaves. 

\begin{figure}
\includegraphics[width=0.7\columnwidth]{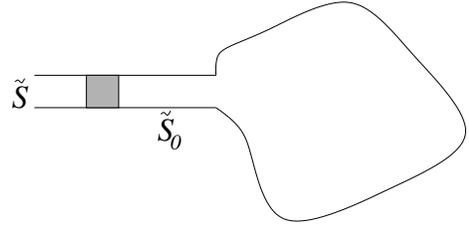} 
\caption{Sketch of a flat chaotic cavity. ${\tilde S}_0$ describes the scattering of the cavity with perfect coupling of the antenna represented as a flat waveguide. ${\tilde S}$ describes the scattering through the system cavity plus a barrier that model the imperfect coupling.}
\label{fig:model}
\end{figure}

When the classical dynamics of the cavity is chaotic, ${\tilde S}$ can be modeled by a random matrix distributed with a probability law. In the presence of both, absorption and  imperfect coupling, the statistical probability distribution of ${\tilde S}$ is given by \cite{Kuhl2004}
\begin{equation}
dP^{(\beta)}_{\langle {\tilde S}\rangle}({\tilde S}) = 
p^{(\beta)}_{\langle {\tilde S}\rangle}({\tilde S})\, dR\, \frac{d\theta}{2\pi}
\label{eq:dP}
\end{equation}
where $\beta=1$ (2) denotes the presence (absence) of time reversal symmetry. Here, 
\begin{equation}
p^{(\beta)}_{\langle {\tilde S}\rangle}({\tilde S}) = 
\left(
\frac{1-\langle {\tilde S}\rangle^2}
{\left|
1 - {\tilde S}\langle S\rangle
\right|^2}
\right)^2
p^{(\beta)}_0(R_0({\tilde S}))
\label{eq:kernel2}
\end{equation}
and $p^{(\beta)}_0(R_0)$ gives the probability distribution 
\begin{equation}
dP^{(\beta)}_0({\tilde S}_0)=p^{(\beta)}_0(R_0)\, 
dR_0\,\frac{d\theta_0}{2\pi}
\label{eq:dP0}
\end{equation}
of the $1\times 1$ scattering matrix 
${\tilde S}_0=\sqrt{R_0}\,e^{i\theta_0}$. This matrix describes the scattering of the system with losses but perfect coupling (see Fig. \ref{fig:model}) . The parameters of ${\tilde S}_0$ are the reflection coefficient $R_0$ and the phase $\theta_0$, in an equivalent way as ${\tilde S}$. Note that $\theta_0$ in Eq. (\ref{eq:dP0}) is uniformly distributed between 0 and $2\pi$, while $R_0$ is distributed according to $p^{(\beta)}_0(R_0)$. For $\beta=2$ \cite{Beenakker2001}
\begin{eqnarray}
p^{(2)}_0(R_0) & = & 
\frac{e^{-\gamma/(1-R_0)}}{(1-R_0)^3} 
\nonumber \\ & \times &
\left[
\gamma\left( e^{\gamma} - 1\right) + 
\left( 1 + \gamma - e^{\gamma} \right)(1-R_0)
\right] \qquad
\label{eq:pb-2}
\end{eqnarray}
while for any $\beta$ (we extend the existing result for $\beta=1$ 
\cite{Savin2005}) 
\begin{equation}
\label{eq:p0(R0)-b1}
p^{(\beta)}_0(R_0) = \frac 2{ (1-R_0)^2 }
P^{(\beta)}_0\left(\frac{1+R_0}{1-R_0}\right),
\end{equation}
where $P^{(\beta)}_0(x)$, with $x=(1+R_0)/(1-R_0)$, leads to the integrated probability distribution 
\begin{equation}
\label{eq:W(x)}
W_{\beta}(x)= \int_x^{\infty}dx\, P^{(\beta)}_0(x).
\end{equation}
$W_{\beta}(x)$ is the positive monotonically decaying function 
\begin{equation}
\label{eq:W(x)-beta2}
W_2(x) = 
\frac 12 
e^{-\gamma x/2}
\left[ e^{\gamma/2}(x+1)-e^{-\gamma/2} (x-1) \right] 
\end{equation}
for $\beta=2$ while for $\beta=1$  
\begin{eqnarray}
\label{eq:W(x)-2}
W_1(x) & = & \frac{x+1}{4\pi}\Big[
f_1(w)g_2(w)+f_2(w)g_1(w) \nonumber \\ 
& + & 
h_1(w)j_2(w)+h_2(w)j_1(w)
\Big]_{w=(x-1)/2},\quad
\end{eqnarray}
with 
\begin{eqnarray}
f_{1(2)}(w) & = & \int_{w(0)}^{\infty(w)}dt
\frac{\sqrt{t|t-w|}e^{-\gamma t/2}}{(1+t)^{3/2}}
\left[1-e^{-\gamma} + t^{-1}\right], \nonumber \\ 
g_{1(2)}(w) & = & \int_{w(0)}^{\infty(w)} dt
\frac 1{\sqrt{t|t-w|}}
\frac{e^{-\gamma t/2}}{(1+t)^{3/2}}, \nonumber \\
h_{1(2)}(w) & = & \int_{w(0)}^{\infty(w)} dt
\frac{\sqrt{|t-w|}e^{-\gamma t/2}}{\sqrt{t(1+t)}} , \label{eq:funcs} \\
&& \quad \qquad \times [\gamma + (1-e^{-\gamma})(\gamma t - 2)] , \nonumber \\
j_{1(2)}(w) & = & \int_{w(0)}^{\infty(w)} dt 
\frac 1{\sqrt{t|t-w|}}
\frac{e^{-\gamma t/2}}{\sqrt{1+t}}.
\nonumber
\end{eqnarray}
Since $W_1(x)$ is very complicated, there are several attempts to interpolate $p^{(1)}_0(R_0)$ between the two well known limits of strong ($\gamma\rightarrow\infty$) and weak ($\gamma\rightarrow 0$) absorption \cite{interpolation,Fyodorov2005}.

The relation between ${\tilde S}$ and ${\tilde S}_0$ is given by the transformation \cite{Kuhl2004}
\begin{equation}
{\tilde S}({\tilde S}_0) = 
\frac{{\tilde S}_0 + \langle {\tilde S}\rangle}
{1+\langle {\tilde S}\rangle {\tilde S}_0},
\label{eq:S(S0)}
\end{equation}
where $\langle{\tilde S}\rangle$ is the averge of ${\tilde S}$. In general $\langle{\tilde S}\rangle$, known as the {\em optical} scattering matrix, is a measure of the prompt responses in the system due to direct processes. In our case an imperfect coupling give rise to direct reflections. The coupling of the antenna to the cavity is quantified by 
\begin{equation}
\label{eq:Ta}
T_a = 1 - |\langle{\tilde S}\rangle|^2
\end{equation}
such that perfect coupling means $\langle{\tilde S}\rangle=0$ or $T_a=1$, and ${\tilde S}$ reduces to ${\tilde S}_0$ as well as $p_{\langle{\tilde S}\rangle}^{(\beta)}({\tilde S})$ becomes the same as $p^{(\beta)}_0({\tilde S}_0)$. Then, by experimental measumerement of $\langle{\tilde S}\rangle$ we can calculate the coupling $T_a$. 


\section{The absorption strength $\gamma$ as a function of 
$\langle R\rangle$ and $T_a$}
\label{sec:gamma}

The average $\langle R\rangle_{\beta}$ can be calculated using directly Eq. (\ref{eq:dP}). However, it is necesary to write $R_0$ as a function of $R$ and $\theta$ (see Eq. (\ref{eq:kernel2})), substitute the resulting expression in the corresponding distribution $p^{(\beta)}_0(R_0)$ above, and integrate with respect to $R$ and $\theta$. Because the resulting expression is still more complicated than 
$p^{(\beta)}_{\langle{\tilde S}\rangle}(R)$, we prefer to calculate 
$\langle R\rangle_{\beta}$ taking advantage of 
$dP^{(\beta)}_{\langle{\tilde S}\rangle}({\tilde S})=
dP^{(\beta)}_0({\tilde S}_0)$ 
and integrate with respect to $R_0$ and $\theta_0$. 

From Eq. (\ref{eq:dP0}) we have 
\begin{equation}
\langle R\rangle_{\beta} = \int_0^{2\pi} \frac{d\theta_0}{2\pi} 
\int_0^1 dR_0\, R(R_0,\theta_0)\, p^{(\beta)}_0(R_0),
\label{eq:aveR-1}
\end{equation} 
where it remains to write $R$ as a function of $R_0$ and $\theta_0$. Using Eq. (\ref{eq:S(S0)}) and that $\langle{\tilde S}\rangle=\sqrt{1-T_a}$ (the phase in the single mode case is not important) we arrive to 
\begin{equation}
\label{eq:R(S0)}
R=
\frac{R_0+(1-T_a)-2\sqrt{1-T_a}\sqrt{R_0}\cos\theta_0}
{1+(1-T_a)R_0-2\sqrt{1-T_a}\sqrt{R_0}\cos\theta_0}.
\end{equation}
If we subtitute this expression for $R$ into Eq. (\ref{eq:aveR-1}) and integrate with respect to $\theta_0$, the result is
\begin{equation}
\label{eq:aveR-2}
\langle R\rangle_{\beta} = 1 - T_a 
\int_0^1 \frac{1-R_0}{1-(1-T_a)R_0} \, 
p^{(\beta)}_0(R_0) \, dR_0,
\end{equation}
where we use that $p^{(\beta)}_0(R_0)$ is normalized to unity. 

We can check the two limits of no coupling and perfect coupling of the antenna to the cavity. For $T_a=0$, $\langle R\rangle_{\beta}=1$ which is compatible with the argument that the waveguide is blocked and the wave never enters the cavity; hence there are no losses. On the opposite, $T_a=1$ leads to 
$\langle R\rangle_{\beta}=\langle R_0\rangle_{\beta}$, which again is expected because the coupling is perfect such that ${\tilde S}={\tilde S}_0$. 

When $0<T_a<1$, the integral in Eq. (\ref{eq:aveR-2}) is not easy to do analytically but it can be done numerically. For $\beta=2$ we directly substitute Eq. (\ref{eq:pb-2}) into Eq. (\ref{eq:aveR-2}) and integrate numerically with respect to $R_0$. The calculation for $\beta=1$ is more complicated. However, a simpler formula for $\langle R\rangle_{\beta}$ for any $\beta$ can be obtained by an integration of Eq. (\ref{eq:aveR-2}) by parts, giving the result (see App. \ref{sec:app1})
\begin{equation}
\langle R\rangle_{\beta} = 1 - T_a + 
2T_a^2 \int_1^{\infty} 
\frac{W_{\beta}(x)}{\left[x+1-(1-T_a)(x-1)\right]^2}dx.
\label{eq:aveR-3}
\end{equation}
The remaining integrations can be done numerically. 

\begin{figure}
\includegraphics[width=0.9\columnwidth]{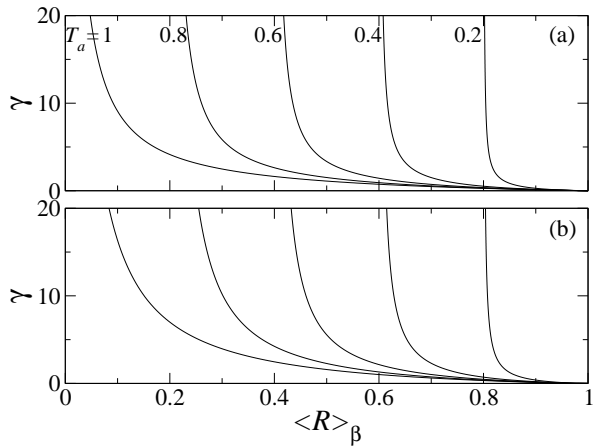} 
\caption{The average of $R$ is a monotonically decreasing function of the losses $\gamma$. We show $\gamma$ as a function of 
$\langle R\rangle_{\beta}$ for several values of the coupling $T_a$ for (a) $\beta=2$ and (b) $\beta=1$. To a given value of $\langle R\rangle_{\beta}$ it corresponds an infinite values of $\gamma$, each for a $T_a$ given.}
\label{fig:avebeta}
\end{figure}

In Fig. \ref{fig:avebeta} we show the results for (a) $\beta=2$ and (b) $\beta=1$ for several values of $T_a$, as can be more useful to experimentalists. We observe that to a single value of $\langle R\rangle_{\beta}$ corresponds an infinite number of $\gamma$ values, each one corresponding to one value of $T_a$. This means that the coupling has a similar effect as the losses on the scattering properties, i. e. the coupling mimic the absorption and viceversa \cite{Mendez-Sanchez2003}. This can be easily seen by writing the argument of the integral of Eq. (\ref{eq:aveR-2}) in powers of $(1-T_a)R_0$ and performing the integral term by term. The result is a useful expression for 
$\langle R\rangle_{\beta}$, namely
\begin{equation}
\langle R\rangle_{\beta} = 1 - T_a + 
T_a^2 \sum_{n=1}^{\infty} (1-T_a)^{(n-1)} 
\langle R_0^n\rangle_{\beta}, 
\label{eq:aveRseries}
\end{equation}
where 
\begin{equation}
\label{eq:aveR0n}
\langle R^n_0\rangle_{\beta} = \int_0^1 R^n_0\, 
p^{(\beta)}_0(R_0) \, dR_0.
\end{equation}
Again, $T_a=0$ implies $\langle R\rangle_{\beta}=1$, while 
$\langle R\rangle_{\beta}=\langle R_0\rangle_{\beta}$ for $T_a=1$. The first two terms in Eq.~(\ref{eq:aveRseries}), $1-T_a$, is the direct reflection due to the coupling at the entrance to the cavity, while the remaining terms are the reflections after the multiple scattering occurs inside the cavity where losses are present. This becomes evident once we calculate below 
$\langle R\rangle_{\beta}$ in both weak and strong absorption regimes.

Then, in an actual experiment it is necessary to measure both parameters, $T_a$ and $\gamma$. Once $T_a$ is calculated from Eq. (\ref{eq:Ta}), $\gamma$ is obtained from $\langle R\rangle_{\beta}$, Eqs.~(\ref{eq:aveR-2}) and (\ref{eq:aveR-3}), which is also obtained from the experimental data. Our formula allows to discern what part of the wave is reflected and what part is lost by absorption. 


\subsection{Strong absorption regime}
\label{subsec:strong}

In this limit, the probability density distribution of $R_0$ reduces to \cite{interpolation,Beenakker2001} 
\begin{equation}
\label{eq:p-strong}
p_0^{(\beta)}(R_0) = 
\alpha_{\beta} 
\frac{e^{-\alpha_{\beta}R_0/(1-R_0)}}{(1-R_0)^{2+\beta/2}}, 
\end{equation}
where we have introduced $\alpha_{\beta}=\gamma\beta/2$. 

Substituting Eq.~(\ref{eq:p-strong}) into Eq.~(\ref{eq:aveR0n}) it is easy to show that (see App. \ref{sec:app2-strong}) 
\begin{equation}
\label{eq:aveR0n-strong}
\langle R_0^n\rangle_{\beta} \approx  
\frac{n!}{\alpha_{\beta}^n} \longrightarrow 0
\quad \mbox{as}\quad 
\alpha_{\beta}\rightarrow\infty,
\end{equation}
which is consistent with the result 
$\langle R_0\rangle_{\beta}=1/\alpha_{\beta}$ of Ref. \cite{Kogan}. 
Then, Eq.~(\ref{eq:aveRseries}) gives 
\begin{equation}
\label{eq:aveR-strong}
\langle R\rangle_{\beta} \approx 1 - T_a.
\end{equation}
This means, the wave entering into the cavity with strong absorption never back and the only reflection occurs at the entrance to the cavity. 


\subsection{Weak absorption regime}
\label{subsec:weak}

In this regime \cite{Beenakker2001},
\begin{equation}
\label{eq:p-weak}
p^{(\beta)}_0(R_0) = 
\frac{\alpha_{\beta}^{1+\beta/2}}{\Gamma(1+\beta/2)} 
\frac{e^{-\alpha_{\beta}/(1-R_0)}}{(1-R_0)^{2+\beta/2}}, 
\end{equation}
where $\Gamma(x)$ is the Gamma function \cite{Abramowitz}. Substituting into Eq.~(\ref{eq:aveR0n}), after some simplifications, we get (see App. \ref{sec:app2-weak})
\begin{equation}
\label{eq:app-weak-aveR0-3} 
\langle R_0^n\rangle_{\beta} \approx 
1 - \frac{2n}{\beta}\alpha_{\beta} =
1 - n\gamma,
\end{equation}
that we insert into Eq. (\ref{eq:aveRseries}) to have
\begin{eqnarray}
\label{eq:aveR-weak2}
\langle R\rangle_{\beta} & \approx & 
1 - T_a - \frac{T_a^2}{1-T_a} +
\frac{T_a^2}{1-T_a} \sum_{n=0}^{\infty} (1-T_a)^n
\nonumber \\ & - &
\gamma\, T_a^2 
\sum_{n=1}^{\infty} n(1-T_a)^{n-1}. 
\end{eqnarray}
The fourth term on the right hand side is just the infinite geometric series and the fifth one its derivative with respect to $(1-T_a)$. They can be summed to give  
\begin{equation}
\label{eq:aveR-weak3}
\langle R\rangle_{\beta} \approx 
1 - \gamma. 
\end{equation}
In this limit a small part of the wave that enters into the cavity is lost by absorption, and the reflection is lightly less than unity.  

\begin{figure}
\includegraphics[width=0.9\columnwidth]{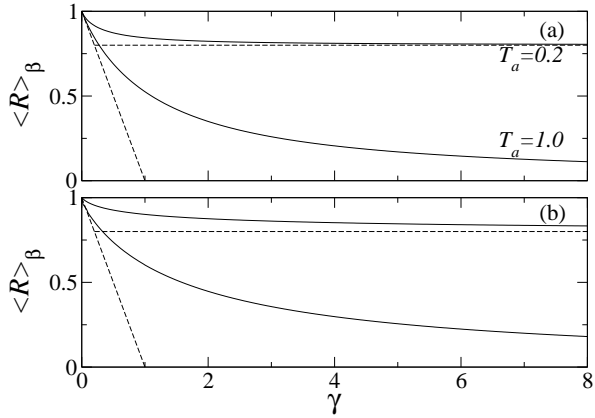} 
\caption{The weak and strong absorption limits of the average of $R$ intersect at $\gamma=T_a$, fixing the criterion for low or high strength of absorption: (a) $\beta=2$ and (b) $\beta=1$.}
\label{fig:transTa10}
\end{figure}

Eqs. (\ref{eq:aveR-strong}) and (\ref{eq:aveR-weak3}) represent the average of $R$ in both limits of strong and weak absorption, respectively, where they are independent of the symmetry $\beta$. The intersection of both limits occurs at $\gamma=T_a$ (see Fig. \ref{fig:transTa10}). Then, the criterion for strong or weak absorption becomes $\gamma\gg T_a$ or $\gamma\ll T_a$ when the coupling is not perfect.

\section{Conclusions}
\label{sec:conclusions}

We have presented a semi-analytical formula to calculate the absorption strength $\gamma$, due to losses in a chaotic cavity, taking the coupling of the single channel port into account. Our result is useful in experiments with microwave networks and one port cavities. We have shown that the imperfect coupling of the antenna to the chaotic systems has a similar effect as the absorption, and viceversa, on the scattering properties. This formula responses to the necesity to calculate an accurate value of $\gamma$. 
We recall that the procedure one channel case can be applied for higher number of channels. 


\appendix

\section{calculation of Eq. (\ref{eq:aveR-3})}
\label{sec:app1}

We start with the substitution of $p_0^{(\beta)}(R_0)$ of Eq. (\ref{eq:p0(R0)-b1}) into Eq. (\ref{eq:aveR-2}). Making the appropriate change of variables, the result is writen as 
\begin{equation}
\label{eq:R1I}
\langle R\rangle_{\beta} = 1 - 2T_a\, I_{\beta}, 
\end{equation}
where
\begin{equation}
I_{\beta} = \int_1^{\infty} A(x) P^{(\beta)}_0(x) dx, 
\end{equation}
with $P^{(\beta)}_0(x)$ giving rise to $W_{\beta}(x)$ as in Eq. (\ref{eq:W(x)}) and 
\begin{equation}
\label{eq:Ax}
A(x) = \frac{1}{x+1-(1-T_a)(x-1)} 
\end{equation}
We integrate by parts identifying 
\begin{eqnarray}
u = A(x), & & du = \frac{dA(x)}{dx} dx, \\
v = -W_{\beta}(x), & & dv = P^{(\beta)}_0(x)dx.
\end{eqnarray}
The result of the integration is
\begin{equation}
\label{eq:I}
I_{\beta} = 
-\left. A(x)W_{\beta}(x)\right|^{\infty}_{x=1} 
+\int_1^{\infty} W_{\beta}(x) \frac{dA(x)}{dx}dx, 
\end{equation}

For $\beta=2$, Eqs. (\ref{eq:W(x)-beta2}) and (\ref{eq:Ax}) gives 
\begin{equation}
\label{eq:eval-beta2}
\left. A(x)W_2(x)\right|^{\infty}_{x=1} = \frac 12.
\end{equation}
We show below that the same result is valid for $\beta=1$.

First, we evaluate $A(x)W_1(x)$ at $x=\infty$, or $w=\infty$, using Eqs.~(\ref{eq:W(x)-2}) and (\ref{eq:Ax}). For instance, we consider the term $f_1(w)g_2(w)$ appearing in Eq.~(\ref{eq:W(x)-2}). Defining $y=t/w$ we can write this term as
\begin{eqnarray}
\label{eq:f1g2oo}
&& f_1(w)g_2(w)|_{w=\infty} 
\nonumber \\ & = & 
(1-e^{-\gamma})
\lim_{w\rightarrow\infty}
\int_1^{\infty}\sqrt{\frac{y(y-1)}{(1+wy)^3}}
we^{-\gamma wy/2} dy 
\nonumber \\ && \times  
\int_0^1 
\frac{we^{-\gamma wy/2}\, \, dy}
{\sqrt{y(1-y)(1+wy)^3}} 
\nonumber \\ & + &
\lim_{w\rightarrow\infty}
\int_1^{\infty}\sqrt{\frac{y-1}{y(1+wy)^3}}
we^{-\gamma wy/2} dy 
\nonumber \\ &&\times
\int_0^1 \frac{e^{-\gamma wy/2}}
{\sqrt{y(1-y)(1+wy)^3}} dy.
\end{eqnarray}
Here, we make use of a definition of the Dirac delta function, namely
\begin{equation}
\lim_{w\rightarrow\infty} we^{-\gamma wy/2} = 
\frac 2{\gamma}\delta(y).
\end{equation}
The interval of integration in Eq.~(\ref{eq:f1g2oo}) does not include the argument of $\delta(y)$; as a consequence $f_1(w)g_2(w)|_{w=\infty}=0$. In a similar way it can be shown that the remaining terms in Eq. (\ref{eq:W(x)-2}) gives zero when evaluated at $w=\infty$. Then, 
\begin{equation}
\label{eq:eval-beta1-1}
A(x)W_1(x)|_{x=\infty} = 0.
\end{equation}

Now, we consider the evaluation of $A(x)W_1(x)$ at $x=1$, or $w=0$.
From Eqs. (\ref{eq:funcs}) it is easy to see that the first term in $W_1(x)$ gives 
\begin{eqnarray}
&& f_1(w)g_2(w)|_{w=0}= \lim_{w\rightarrow 0} \int_0^1 
\frac{e^{-\gamma wy/2}}{\sqrt{y(1-y)(1+wy)^3}} dy
\nonumber \\ & &\times 
\int_w^{\infty}
\sqrt{\frac{t(t-w)}{(1+t)^3}}e^{-\gamma t/2}
\left( 1 - e^{-\gamma} + \frac 1t\right)dt, 
\end{eqnarray}
which reduces to
\begin{equation}
\label{eq:f1g2}
f_1(w)g_2(w)|_{w=0}= \pi 
\int_0^{\infty}
\frac{te^{-\gamma t/2}}{(1+t)^{3/2}}
\left( 1 - e^{-\gamma} + \frac 1t\right)dt. 
\end{equation}
Similarly, we may also shown that the third term in $W_1(x)$ gives 
\begin{eqnarray}
\label{eq:h1j2}
h_1(w)j_2(w)|_{w=0} & = & \pi 
\int_0^{\infty}
\frac{e^{-\gamma t/2}}{\sqrt{1+t}} 
\nonumber \\ &\times &
\left[ \gamma + \left(1 - e^{-\gamma}\right)
\left(\gamma t - 2\right)\right]dt. \qquad\quad
\end{eqnarray}

The second and the fourth terms in $W_1(x)$ are zero when evaluated at $x=1$. We consider the second one only. It is 
\begin{eqnarray}
&& f_2(w)g_1(w)|_{w=0} 
\nonumber \\ & = &
\lim_{w\rightarrow 0} \int_0^1 
\sqrt{\frac{y(1-y)}{(1+wy)^3}} w^2e^{-\gamma wy/2} 
\left( 1 - e^{-\gamma} + \frac 1{wy}\right)dy
\nonumber \\ & \times & 
\int_w^{\infty} 
\frac{e^{-\gamma t/2}} {\sqrt{t(t-w)(1+t)^3}}dt, \qquad\quad
\end{eqnarray}
which reduces to zero. From Eqs.~(\ref{eq:f1g2}), (\ref{eq:h1j2}), and  (\ref{eq:Ax}) we get
\begin{eqnarray}
\label{eq:A0W0part}
&& A(x)W_1(x)|_{x=1} = 
\frac 14 (1-e^{-\gamma})\int_0^{\infty}
\frac{te^{-\gamma t/2}}{(1+t)^{3/2}}
\nonumber \\ && +
\int_0^{\infty}
\frac{e^{-\gamma t/2}}{(1+t)^{3/2}} + 
(\gamma - 2 + 2e^{-\gamma})\int_0^{\infty}
\frac{e^{-\gamma t/2}}{(1+t)^{1/2}}
\nonumber \\ && +
\gamma (1 - e^{-\gamma})
\int_0^{\infty}
\frac{te^{-\gamma t/2}}{(1+t)^{1/2}}.
\end{eqnarray}

Integrating by parts, we can establish the following relations
\begin{eqnarray}
\int_0^{\infty} \frac{e^{-\gamma t/2}}{(1+t)^{1/2}}dt & = & 
\frac 2{\gamma} - 
\frac 1{\gamma}\int_0^{\infty} \frac{e^{-\gamma t/2}}{(1+t)^{3/2}}dt, 
\label{eq:rel1}
\\ 
\int_0^{\infty} \frac{te^{-\gamma t/2}}{(1+t)^{1/2}}dt & = & 
\left(\frac 2{\gamma}\right)^2 - 
\frac 1{\gamma}\int_0^{\infty} \frac{te^{-\gamma t/2}}{(1+t)^{3/2}}dt 
\nonumber \\ & - & \frac2{\gamma^2} 
\int_0^{\infty} \frac{e^{-\gamma t/2}}{(1+t)^{3/2}}dt.
\label{eq:rel2}
\end{eqnarray}
Substituting Eqs.~(\ref{eq:rel1}) and (\ref{eq:rel2}) into Eq.  (\ref{eq:A0W0part}) we obtain
\begin{equation}
\label{eq:eval-beta1-2}
A(x)W_1(x)|_{x=1} = \frac 12.
\end{equation}
Therefore, using Eqs. (\ref{eq:eval-beta2}), (\ref{eq:eval-beta1-1}), and (\ref{eq:eval-beta1-2}), Eq. (\ref{eq:I}) can be written as 
\begin{equation}
\label{eq:Ifinal}
I_{\beta} = \frac 12 + \int_1^{\infty}W_{\beta}(x)\frac{dA(x)}{dx}dx, 
\end{equation}
Finally, Eqs.~(\ref{eq:R1I}), (\ref{eq:Ax}), and (\ref{eq:Ifinal}) gives the result shown in Eq. (\ref{eq:aveR-3}).


\section{Calculation of $\langle R_0^n\rangle_{\beta}$}

\subsection{Strong absorption limit}
\label{sec:app2-strong}

In this limit, Eq.~(\ref{eq:p-strong}) can still be simplified to the Rayleigh distribution \cite{interpolation,Kogan}
\begin{equation}
\label{eq:Rayleigh}
p_0^{(\beta)}(R_0) = 
\alpha_{\beta} \,
e^{-\alpha_{\beta}R_0}, 
\end{equation}
which substituted into Eq.~(\ref{eq:aveR0n}) lead us to 
\begin{eqnarray}
\langle R_0^n\rangle_{\beta} & = & \alpha_{\beta}
\int_0^1R_0^n\, e^{-\alpha_{\beta} R_0} dR_0 
\nonumber \\ & = &
-e^{-\alpha_{\beta}} + \frac n{\alpha_{\beta}}
\langle R_0^{n-1}\rangle_{\beta},
\end{eqnarray}
where an integration by parts was done. This expression can be iterated to  obtain the precise result  
\begin{equation}
\langle R_0^n\rangle_{\beta} = 
\frac{n!}{\alpha_{\beta}^n} - 
e^{-\alpha_{\beta}} \sum_{m=0}^{n-1} 
\frac{n!}{(n-m)!\alpha_{\beta}^m},
\end{equation}
where the limit $\alpha_{\beta}\rightarrow\infty$ gives the result shown in Eq.~(\ref{eq:aveR0n-strong}).


\subsection{Weak absorption limit}
\label{sec:app2-weak}

We substitute Eq.~(\ref{eq:p-weak}) into Eq.~(\ref{eq:aveR0n}) to have 
\begin{eqnarray}
\langle R_0^n\rangle_{\beta} & = & 
\frac{\alpha_{\beta}^{1+\beta/2}}
{\Gamma(1+\beta/2)}
\int_0^1 R_0^n 
\frac{e^{-\alpha_{\beta}/(1-R_0)}}{(1-R_0)^{2+\beta/2}}
dR_0 
\nonumber \\ 
\label{eq:app-weak-aveR0-1} 
& = & 
\frac{\alpha_{\beta}^{1+\beta/2}}
{\Gamma(1+\beta/2)}
\int_1^{\infty} e^{-\alpha_{\beta}x}
(x-1)^n x^{\frac{\beta}2-n} dx, \qquad
\end{eqnarray}
where we have change to the variable $x=(1-R_0)^{-1}$. 
Using the binomial expansion for $(x-1)^n$ we write Eq.~(\ref{eq:app-weak-aveR0-1}) as 
\begin{equation}
\label{eq:app-weak-aveR0-2} 
\langle R_0^n\rangle_{\beta} = 
\frac 1{\Gamma(1+\beta/2)}
\sum_{r=0}^n 
\frac{(-\alpha_{\beta})^{n-r}n!}{r!(n-r)!} 
\Gamma(1+\beta/2+r-n,\alpha)
\end{equation}
where $\Gamma(a,x)$ is the incomplete Gamma function \cite{Abramowitz}. Keeping linear terms in $\alpha_{\beta}$ only, we arrive to the result shown in Eq.~(\ref{eq:app-weak-aveR0-3}). We recall that special care for $\beta=2$ 
should be taken because of the divergence of the Gamma functions, however  result in Eq.~(\ref{eq:app-weak-aveR0-3}) is valid.


\acknowledgments

The authors thank financial support from DGAPA-UNAM, M\'exico, under project IN118805.



\end{document}